\documentclass[twocolumn,aps,pra,showpacs,floatfix]{revtex4}
\usepackage{graphicx}
\usepackage{times}
\usepackage{nicefrac}
\usepackage{amsmath}
\usepackage{amsfonts}
\usepackage{amssymb}
\usepackage{amsthm}
\usepackage{epsf}
\usepackage{bm}
\usepackage{bbm}

\usepackage{dcolumn}
\newcolumntype{.}{D{x}{}{-1}}

\newcommand{\vare}{\varepsilon}
\newcommand{\Za}{Z\alpha}

\newcommand{\bgamma}{\bm{\gamma}}

\newcommand{\bfp}{\bm{p}}

\newcommand{\bfx}{\bm{x}}

\newcommand{\pr}{^{\prime}}

\begin{document}

\title{The two-loop self-energy for the ground state of medium-$\bm{Z}$ hydrogen-like ions}

\author{V.~A.~Yerokhin}
\affiliation{Center for Advanced Studies, St.~Petersburg State
Polytechnical University, Polytekhnicheskaya 29, 
St.~Petersburg 195251, Russia}

\begin{abstract}
The two-loop self-energy correction to the ground state Lamb shift is
calculated for hydrogen-like ions with the nuclear charge $Z=10\ldots30$ without any
expansion in the binding field of the nucleus. A calculational technique is
reported for treatment of Feynman diagrams in the mixed coordinate-momentum
representation, which yields significant improvement in numerical accuracy as compared to
previous results. An extrapolation of the all-order numerical data towards $Z=1$ 
yields a result for the higher-order remainder function for hydrogen.
The previously reported disagreement between the all-order and the
perturbative approaches is reduced to the marginal agreement. 
\end{abstract}


\maketitle

Calculations of the two-loop QED corrections to all orders in the 
binding-strength parameter of the nucleus $\Za$ ($Z$ is the nuclear charge and
$\alpha$ is the fine structure constant) were motivated almost two decades ago by 
progress in the experimental spectroscopy of heavy lithium-like ions
\cite{schweppe:91}. To date, experimental investigations of the $2p_J$-$2s$
transitions in high- and medium-$Z$ lithium-like ions are sensitive to the two-loop QED
effects on the level of about 10\% \cite{beiersdorfer:05,epp:prl:07}.  
All-order calculations of the two-loop
Lamb shift were recently accomplished in Ref.~\cite{yerokhin:03:prl} for high-$Z$ hydrogen-like
ions and in Ref.~\cite{yerokhin:06:prl} for high-$Z$ lithium-like ions. 

The ratio of the QED effects to the binding energy scales as $Z^2$,
so that the relative QED contribution gets smaller for lighter ions. However,
the experimental precision is also better there and the two-loop QED effects
in light systems have long being observed.
The best studied case is atomic hydrogen, whose spectroscopy is nowadays
carried out with an accuracy of a few parts in $10^{14}$ \cite{niering:00}. 

Until recently, calculations of the QED effects in hydrogen relied on the approach
perturbative in the binding-strength parameter $\Za$. Technical difficulties
of this approach, however, grow rapidly with the increase of the order of the
perturbative expansion. The state of the art of such calculations is the
evaluation of the dominant part of the $\alpha^2\,(\Za)^6$
correction \cite{pachucki:01:pra,pachucki:03:prl,jentschura:05:sese}. 
An alternative way is to perform a numerical calculation to all orders in
$\Za$ and to identify the higher-order remainder by subtracting
the known low-order terms from the all-order results. The problems on this way
are, first, significant internal cancellations in numerical calculations, 
which grow as $Z$ decreases, and second, additional losses of accuracy
occuring when the higher-order remainder is inferred from the all-order results. 

The first attempt at the evaluation of the all-order remainder for the ground state
of hydrogen was made in Ref.~\cite{yerokhin:05:sese}. In that work, a
numerical calculation of the two-loop self-energy correction was reported for
$Z\ge 10$. This correction is expected to give the dominant contribution to
the two-loop remainder. An extrapolation towards $Z=1$ performed
in Ref.~\cite{yerokhin:05:sese} yielded a result for the higher-order
remainder approximately twice as large as the estimate based on the analytical calculations
\cite{pachucki:03:prl}. This disagreement is presently the main source of the
theoretical uncertainty of the ground-state Lamb shift in hydrogen and
influences the values of the Rydberg constant and the proton charge radius
obtained from the hydrogen spectroscopic data \cite{mohr:08:rmp}. 

This investigation presents an attempt to resolve the disagreement
between the numerical and analytical approaches by
improving the calculational accuracy of the all-order results. To this end, we
develop a scheme for the evaluation of Feynman diagrams in the mixed
coordinate-momentum representation (the corresponding part of the two-loop
self-energy is conventionally termed as the $P$ term). This is the most
nontrivial part of the evaluation of the two-loop self-energy correction, as it has no
analog in the one-loop calculations. 

For the first time the $P$ term was calculated in
Ref.~\cite{yerokhin:01:sese} with help of a finite basis set representation of the
spectrum of the Dirac equation. Later investigations 
\cite{yerokhin:03:prl,yerokhin:05:sese,yerokhin:06:prl} proceeded along the
same way with adopting an improved (dual kinetically ballanced) basis set
\cite{shabaev:04:DKB}. The main problem of the basis-set approach is
a relatively slow convergence  with respect to the number of basis functions. 
In order to overcome this limitation, in this work we employ the analytical
representation of the Dirac-Coulomb Green function (DCGF) in terms of the Whittaker
functions. As a result, the numerical accuracy of the $P$
term is improved by more than an order of magnitude, the error now being
mainly due to the termination of the partial-wave expansion. 

In the present investigation, we perform a reevaluation of the two-loop
self-energy correction for the ground state of hydrogen-like ions with the nuclear
charge numbers in the interval $Z = 10\ldots30$. The calculation of the $P$
term is carried out with the technique developed in this work. The other parts
of the correction are evaluated by the methods described previously
\cite{yerokhin:03:epjd} but with the increased number of partial waves
included and with denser integration grids. The nonperturbative remainder
incorporating terms of order $\alpha^2\,(\Za)^6$ and higher is inferred from
the numerical results and extrapolated toward $Z=1$. 

\begin{figure}
\begin{center}\includegraphics[width=\columnwidth]{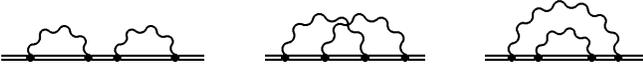}\end{center}%
\caption{\label{fig1}
The two-loop self-energy correction. The double line represents the electron
propagator in the Coulomb binding field of the nucleus. }%
\end{figure}

\begin{figure}
\begin{center}\includegraphics[width=\columnwidth]{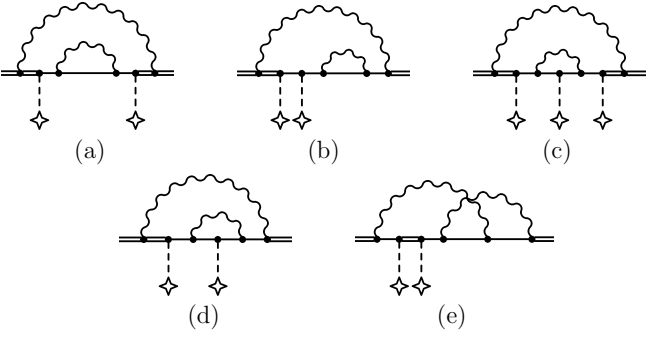}\end{center}%
\caption{\label{fig2}
The $P$ term. The single line represents the free electron propagator. The
dashed line with a cross indicates the interaction with the Coulomb field of
the nucleus.}
\end{figure}

The two-loop self-energy correction is shown in Fig.~\ref{fig1}. In this
work, we concentrate on the part of the correction that is treated in the
mixed coordinate-momentum representation and is referred to as the $P$
term. The corresponding Feynman diagrams are shown in Fig.~\ref{fig2}. They
arise from the diagrams in Fig.~\ref{fig1} when the bound-electron propagators
are expanded in terms of the interaction with the binding
field. The distinct feature of the diagrams contributing to the $P$ term is
that ultraviolet divergences in them originate from the one-loop
subgraphs only. These subgraphs are covariantly regularized and calculated in 
the momentum space, whereas the remaining part of the diagrams does not need
any regularization and is treated in the coordinate space. 


In order to illustrate the calculational technique used for the evaluation of
the $P$ term, we
consider one of the simplest diagrams in Fig.~\ref{fig2}, graph (b). 
Its contribution can be written as 
\begin{align}                                           \label{n1p5}
\Delta E_{N1b,P} &\  = 
        4i\alpha \int_{C_F} d\omega\,
                \int \frac{d\bfp}{(2\pi)^3}\,
                        \int d\bfx_1 d \bfx_4 \,
        D(\omega,x_{14})\,
                \nonumber \\ & \times
                 \psi^{\dag}_a(\bfx_1)\, \alpha_{\mu}
\Bigl[G_V(E,\bfx_1,\bfp)-G_V^{(0)}(E,\bfx_1,\bfp)\Bigr] 
                \nonumber \\ & \times
\frac1{\gamma^0E-\bgamma\cdot \bfp-m}\,  \Sigma_R^{(0)}(E,\bfp)\,
    G^{(0)}(E,\bfp,\bfx_4)\,
                \nonumber \\ & \times
         \alpha^{\mu}\,        \psi_a(\bfx_4)\, ,
\end{align}
where $E = \vare_a-\omega$, $\vare_a$ is the reference-state energy,
$D(\omega,x_{14})$ is the scalar part of the photon propagator in the
Feynman gauge defined by $D^{\mu\nu}(\omega,x_{14}) \equiv g^{\mu\nu}\,D(\omega,x_{14})$,
$\Sigma_R^{(0)}$ is the renormalized free self-energy operator 
(see, e.g., Ref.~\cite{yerokhin:03:epjd} for its definition and evaluation),
$G_V(\vare,\bfx_1,\bfp)$ is the Fourier transform of the product of the
DCGF $G(\vare,\bfx_1,\bfx_2)$ and the Coulomb
potential $V_C$,
\begin{equation} \label{n1p3}
G_V(\vare,\bfx_1,\bfp) = 
        \int d\bfx_2 \, e^{i\bfp\cdot \bfx_2} \,
       G(\vare,\bfx_1,\bfx_2) \, V_C(\bfx_2)
         \,,
\end{equation} 
the function $G_V^{(0)}$ is given by an analogous expression with the
DCGF substituted by the free Dirac Green function $G^{(0)}$,
and $G^{(0)}(E,\bfp,\bfx_4)$ is the Fourier transform of the free Dirac Green
function.

The main problem of the evaluation of Eq.~(\ref{n1p5}) is the absence of
a convenient representation for the function $G_V(\vare,\bfx_1,\bfp)$. 
Because of this, we have to start with the
the coordinate-space representation of the Green function and perform the
Fourier transfomation numerically. The DCGF is
known in terms of the partial-wave expansion over the relativistic angular
parameter $\kappa$, with the radial part given by \cite{mohr:98}
\begin{equation}\label{greena}
 G_{\kappa}(\vare,x_1,x_2) =  
\left\{ 
\begin{array}{ll}
-\phi_{\kappa}^{0}(\vare,x_1)\,\phi_{\kappa}^{{\infty}^T}(\vare,x_2)\,, & \mbox{for}\
           x_1 <x_2\,,\\
-\phi_{\kappa}^{\infty}(\vare,x_1)\,\phi_{\kappa}^{{0}^T}(\vare,x_2)\,, & \mbox{for}\
           x_2 <x_1\,,
\end{array}
\right.
\end{equation}
where $\phi_{\kappa}^{0}$ and $\phi_{\kappa}^{\infty}$ are the two-component
solutions of the radial Dirac equation regular at the origin and at the infinity,
respectively, and normalized in such a way that their Wronskian is unity. The Fourier
transform of the DCGF over the second radial variable can be represented as
\begin{align}\label{greenb}
 G_{\kappa}(\vare,x_1,p) = &\
 -\phi_{\kappa}^{\infty}(\vare,x_1)\,\psi_{\kappa}^{0^T}(\vare,p;x_1)
\nonumber \\ &
 -\phi_{\kappa}^{0}(\vare,x_1)\,\psi_{\kappa}^{{\infty}^T}(\vare,p;x_1)\,,
\end{align}
with
\begin{align}\label{gr3}
\psi_{\kappa}^{0}(\vare,p;x_1) = 4\pi\,\int_0^{x_1}dx_2\,
  x_2^2\, \left( 
        \begin{array}{r}
   j_l(px_2)\,\phi^0_{\kappa,+}(\vare,x_2) \\[0.5em]
   -\frac{\kappa}{|\kappa|\,}j_{\overline{l}}(px_2)\,\phi^0_{\kappa,-}(\vare,x_2) \\
        \end{array}
\right)\,,
\end{align}
and
\begin{align}\label{gr4}
\psi_{\kappa}^{\infty}(\vare,p;x_1) = 4\pi\,\int_{x_1}^{\infty} dx_2\,
  x_2^2\, \left( 
        \begin{array}{r}
   j_l(px_2)\,\phi^{\infty}_{\kappa,+}(\vare,x_2) \\[0.5em]
   -\frac{\kappa}{|\kappa|\,}j_{\overline{l}}(px_2)\,\phi^{\infty}_{\kappa,-}(\vare,x_2) \\
        \end{array}
\right)\,,
\end{align}
where $\phi^{0}_{\kappa,\pm}$ and $\phi^{\infty}_{\kappa,\pm}$ denote the upper and
lower components of $\phi^{0}_{\kappa}$ and $\phi^{\infty}_{\kappa}$,
respectively, $l = |\kappa+1/2|-1/2$ and $\overline{l} = |\kappa-1/2|-1/2$.

The integration over $x_2$ in the functions $\psi_{\kappa}(\vare,p;x_1)$ has to be
performed numerically. The problems here are that (i) the integration interval depends on
$x_1$ and (ii) the integrand contains the spherical Bessel function which 
oscillates rapidly in the high-momenta region. 
Clearly, a straightforward use of Eqs.~(\ref{gr3}) and (\ref{gr4})
would lead to a re-evaluation of the integral for each new value of $x_1$,
making the calculation prohibitively expensive.  
One can observe, however, that if the function $\psi_{\kappa}(\vare,p;x)$ is known
for a particular set of $\vare$, $p$, and $x$, then the evaluation of
$\psi_{\kappa}(\vare,p;x\pr)$ can be done by computing the Bessel
transform integral over the interval $(x,x\pr)$ only. So, introducing an ordered
radial grid $\left\{x_i\right\}$, one can store
the set of values $\left\{\psi_{\kappa}(\vare,p;x_i)\right\}$ by performing just one Bessel
transform over the interval $(0,\infty)$. This shows that for a fixed values of $\vare$ and $p$,
the integration over $x_1$ can be performed without a recalculation of the
Bessel transform integral. Still, the evaluation of the functions $\psi_k$
was one of the most problematic parts of the computation since a controllable
accuracy was required for momenta as high as $p=10^6$. 

The next problem to be solved in the numerical evaluation of Eq.~(\ref{n1p5}) 
is that the free Dirac Green function $G^{(0)}(E,\bfp,\bfx_4)$ contains a 
spherical Bessel function $j_L(px_4)$ and thus is highly oscillating too 
in the high-momenta region. It can be observed that the radial integration 
over $x_4$ resembles the Bessel transform integral over $x_2$ and so can be
efficiently calculated by introducing analogs of the functions
$\psi_k$. Details of the numerical procedure will be published elsewhere.  

The results of our numerical evaluation of the two-loop self-energy correction
to the ground-state Lamb shift of middle-$Z$ hydrogen-like ions 
are presented in Table~\ref{tab:sese}. They are consistent with but improve upon
the data obtained previously \cite{yerokhin:05:sese,yerokhin:05:jetp}.

%%%%%%%%%%%%%%%%%%%%%%%%%%%%%%%%%%%%%%%%%%%%%%%%%%%%%%%%%%%%%%%%%%%%%%%%%%%%%%%%%%%
%
%   Total results for SESE
%
%%%%%%%%%%%%%%%%%%%%%%%%%%%%%%%%%%%%%%%%%%%%%%%%%%%%%%%%%%%%%%%%%%%%%%%%%%%%%%%%%%%
\begin{table*}[htb]
\caption{The two-loop self-energy
correction for the ground state of hydrogen-like ions, in units of $\Delta
E/[m\alpha^2(\Za)^4/\pi^2]$. ``LAL'' denotes the loop-after-loop correction. 
Definitions and detailed description of individual contributions can be found in
Ref.~\cite{yerokhin:03:epjd}. 
 \label{tab:sese}}
\begin{ruledtabular}
\begin{tabular}{r......}
$Z$  &  \multicolumn{1}{c}{LAL}
              &  \multicolumn{1}{c}{$F$ term}
                             &  \multicolumn{1}{c}{$P$ term}
                                      &  \multicolumn{1}{c}{$M$ term}
                                               &  \multicolumn{1}{c}{Total}
                                               &  \multicolumn{1}{c}{2005
                                                 results
                                                 \cite{yerokhin:05:sese}}\\
\hline
10 &  -0.3x58 &  822.x138\,(5) &  -721.x311\,(6) & -100.x297\,(35) &   0.17x2\,(36) & 0.25x\,(16) \\
12 &  -0.4x17 &  519.x603\,(2) &  -439.x065\,(6) &  -80.x117\,(38) &   0.00x4\,(38) & \\
15 &  -0.4x95 &  292.x901\,(2) &  -235.x211\,(4) &  -57.x406\,(11) &  -0.21x2\,(12) & -0.16x4\,(85) \\
17 &  -0.5x41 &  211.x052\,(1) &  -164.x280\,(3) &  -46.x567\,(9)  &  -0.33x6\,(10) & \\
20 &  -0.6x02 &  136.x909\,(1) &  -102.x029\,(2) &  -34.x780\,(4)  &  -0.50x1\,(5)  & -0.48x1\,(58)\\
25 &  -0.6x86 &   74.x501\,(1) &   -51.x982\,(2) &  -22.x560\,(6)  &  -0.72x8\,(6)  & \\
30 &  -0.7x56 &   44.x728\,(1) &   -29.x414\,(3) &  -15.x468\,(3)  &  -0.91x0\,(5)  & -0.90x3\,(26) \\
\end{tabular}
\end{ruledtabular}
\end{table*}

\begin{figure}
\begin{center}\includegraphics[width=\columnwidth]{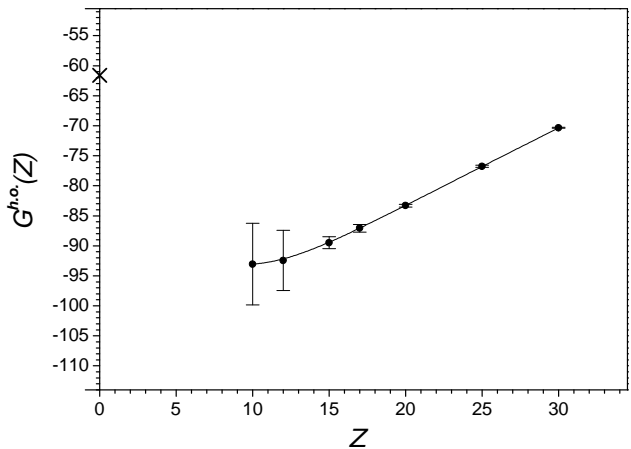}\end{center}%
\caption{\label{fig:ho}
The higher-order remainder function of the two-loop self-energy correction.
The cross on the $y$-axis indicates the analytical result. 
}%
\end{figure}

In the present investigation we are concerned with the higher-order remainder
function that incorporates contributions of all orders starting with
$\alpha^2(\Za)^6$ and is denoted as $G^{\rm h.o.}_{\rm SESE}$. It  is obtained from the 
two-loop self-energy correction $\Delta E_{\rm SESE}$ by separating out the first terms of
its $\Za$ expansion,
\begin{align}
\label{FalphaZ} \Delta E_{\rm SESE} &\ = m \left(\frac{\alpha}{\pi}\right)^2
(Z\alpha)^4\,
 \Bigl\{
B_{40}+
(Z\alpha)B_{50} 
\nonumber \\ &
+ (Z\alpha)^2 \Bigl[
  L^3 B_{63}
  +L^2 B_{62} +  L\,B_{61} + G^{\rm h.o.}_{\rm SESE}(Z) \Bigr]
 \Bigr\}
\,,
\end{align} 
where  $L \equiv \ln[(Z\alpha)^{-2}]$  and the expansion of the remainder 
starts with a constant, $G^{\rm h.o.}_{\rm SESE}(Z) = B_{60}+ \Za \,(\ldots)\,$. The results
for the expansion coefficients (see
Refs.~\cite{pachucki:01:pra,pachucki:03:prl,jentschura:05:sese,mohr:08:rmp} and references
therein) are: $B_{40} = 1.409244$, $B_{50}= -24.2668(31)$, $B_{63}= -8/27$,
$B_{62}= 16/27-(16/9) \ln 2$, $B_{61} = 48.388913$, and $B_{60} = -61.6(9.2)$.
The remainder function inferred from our numerical results is plotted in
Fig.~\ref{fig:ho}. 

%%%%%%%%%%%%%%%%%%%%%%%%%%%%%%%%%%%%%%%%%
%%%%%%%%%%%%%%%%%%%%%%%%%%%%%%%%%%%%%%%%%

In order to obtain a value of the remainder function for
hydrogen, we have to extrapolate the numerical data obtained for $Z\ge 10$
towards $Z=1$. For this we use a variant of the procedure first 
employed in Ref.~\cite{mohr:75:prl}.  The extrapolated
value of $G^{\rm h.o.}_{\rm SESE}(Z=1)$ is obtained in two steps. First, we apply an
(exact) linear fit to each pair of two consecutive points from our data set
and store the resulting values at $Z=1$. Second, we perform a global parabolic 
least-squares fit to the set of data obtained on the first step 
and take the fitted value at
$Z=1$ as a final result. Similar procedure applied to the determination of the
$B_{50}$ coefficient reproduces the known analytical result with the accuracy
of about 1\%. For comparison, a global polynomial fit yields a result
for the $B_{50}$ coefficient accurate within 5\% only. 

When applied to the remainder function
$G^{\rm h.o.}_{\rm SESE}(Z)$, the extrapolation procedure described above gives 
\begin{equation} \label{ho}
G^{\rm h.o.}_{\rm SESE}(Z=1) = -86\,(15)\,.
\end{equation}
This value is significantly higher than what would be expected simply from a
polynomial fitting (of about $-105$), which suggests the presence of unusually
large logarithmic contributions to order $\alpha^2(\Za)^7$ (logarithms of up to the
second power are anticipated \cite{jentschura:05:sese}).  
The value (\ref{ho}) is also higher than but consistent with the 2005 result of $-127\,(42)$ 
\cite{yerokhin:05:sese}. The shift of the central value is due to two
reasons. First, the analytical value of the $B_{61}$ coefficient was changed
during this time by $\delta B_{61}=-1.4494\ldots$ \cite{jentschura:05:sese}, thus
pushing the higher-remainder higher up. Second, the improved numerical
accuracy of the present calculation and the increased number of data points
allowed us to identify the upward trend in the numerical data, which influenced
the extrapolated value considerably. The present result (\ref{ho}) is marginally
consistent with the analytical value of $B_{60} = -62(9)$. 

Finally, we account for the contribution from the
diagrams with closed fermion loops calculated recently in
Ref.~\cite{yerokhin:08:twoloop}. The total value of the remainder function for
hydrogen is
\begin{equation}
G^{\rm h.o.}(Z=1) = -86\,(15)-15\,(2) = -101\,(15)\,.
\end{equation}

To conclude, the present investigation reports a technique for the evaluation
of Feynman diagrams in the mixed coordinate-momentum representation, which
allows one to significantly improve the numerical accuracy. A complete
recalculation of the diagrams of the two-loop self-energy is presented for the
ground state of hydrogen-like ions with the nuclear charge number $Z=10-30$. 
The higher-order (in $\Za$) remainder function is inferred from numerical
all-order results and extrapolated towards $Z=1$. The extrapolated value of
the higher-order remainder function is in marginal agreement with the
analytical result obtained within the perturbative approach. 

The work reported in this paper was supported by the ``Dynasty'' foundation.

%\bibliographystyle{/home/yerokhin/papers/bibtex/phaip30}
%\bibliography{/home/yerokhin/papers/bibtex/hfst}

\end{document}